\documentclass[conference]{IEEEtran}
\usepackage{graphicx} 
\usepackage{caption} 
\usepackage{comment} 
\usepackage{subcaption}
\usepackage{tikz}
\usetikzlibrary{calc}
\usepackage{makecell}
\usetikzlibrary{quantikz}
\usepackage{lscape} 
\usepackage{amsmath} 
\usepackage{braket}  
\usepackage{tabularx} 
\newcolumntype{Y}{>{\centering\arraybackslash}X} 
\usepackage{multirow} 

\title{Security Implications of Measurement Based Crosstalk on Superconducting Quantum Computers\\}
\author{\IEEEauthorblockN{Subarna Adhikari}
\IEEEauthorblockA{\textit{OUSPG} \\
\textit{University of Oulu}\\
Oulu, Finland \\
subarna.adhikari@oulu.fi}
\and
\IEEEauthorblockN{Kimmo Halunen}
\IEEEauthorblockA{\textit{OUSPG} \\
\textit{University of Oulu}\\
Oulu, Finland \\
kimmo.halunen@oulu.fi}
}

\usepackage[backend=biber,style=ieee]{biblatex}
\usepackage{xurl}
\begin{document}

\maketitle
\setlength{\parindent}{0pt}
\begin{abstract} 
    Crosstalk is one of the major issues for scalability of superconducting quantum processors. But it can also be used by threat actors to intentionally sabotage computations and steal information. This paper analyzes data leakage due to crosstalk during qubit measurement. Prior research yielded an accuracy of 96\% in classifying measurement of one qubit in IBM quantum computers. When evaluated on superconducting hardware from VTT, we achieved an accuracy of 71.68\% with similar framework. Our extended framework with additional labels yields an accuracy of 76.30\% for classifying measurement of one qubit and 53.37\% for measurement of two qubits using Support Vector Machine (SVM). Experimental results indicate that  measurement operation on a qubit can affect the probability of measuring adjacent qubits. The impact on the target qubit is dependent on the number of neighboring qubits measured and also the measurement outcome.

\end{abstract}


\section{Introduction}
    \label{introduction}
    Quantum computing is moving towards practical deployment. Studies have explored the use of quantum computing in solving complex problems that can aid business performance and research breakthroughs \cite{GILL20261}  \cite{hassija2020forthcoming}. Some companies are already undertaking experimental projects to examine the potential role of Quantum Computers (QCs) in real-world applications \cite{GILL20261}. However, this also raises concerns regarding the security and trust in QCs. Such considerations are especially crucial with the multi-tenant user model, where users and businesses get shared access to QCs over the cloud. This is further complicated by the fact that there isn't a universal hardware implementation of QC. They differ from each other in qubit representation, control hardware, software processes and environmental factors. As such, research efforts need to consider all architectures before we have a comprehensive understanding of the security of QCs. Such efforts will also enable users and businesses to make informed decisions regarding QCs based on their security needs. Additionally, as seen with classical cybersecurity, identifying security issues in the early stages is likely to reduce the financial overhead associated with mitigation efforts at later stages \cite{kumar2021cost}. Addressing security challenges right from the beginning also facilitates informed design decisions that improve robustness and system trustworthiness.

Most research on security of QCs focus on crosstalk which can cause interference between simultaneous operations \cite{Adhikari2025review}. Crosstalk can arise from multiple sources like qubit interactions, control pulses and electromagnetic interference \cite{sarovar_detecting_2020}. A platform independent definition of crosstalk is provided by the authors in \cite{sarovar_detecting_2020}. They define crosstalk as the violation of locality or independence or both. Locality can be described as the restriction of the quantum operations to the qubits involved whereas, independence signifies that simultaneous operations do not affect each other. Crosstalk has been one of the major challenges for the scalability of current generation of superconducting QCs. However, crosstalk can also be leveraged by attackers to influence or obtain information about the computations of other users in the system \cite{ash2020analysis}\cite{choudhury2024}.

Regarding vulnerabilities related to qubit measurement, the authors in \cite{saki2021qubit} examined the correlation between measurement of adjacent qubits. They observed that measuring a qubit state `0' or `1' and can affect simultaneous measurement of `1' on the neighboring qubit. Our works builds on their methodology by examining this correlation in additional scenarios (measuring `0', `1' and no measurement). All the scenarios and the representative circuits are explained in section IV.

The outputs from executing the circuits were used to construct datasets for predicting the state of the target qubits (victim qubits). The probability of measuring the neighboring qubit (attacker qubit) is a characteristic feature in the dataset. The datasets were used to train and test Support Vector Machine (SVM) classifiers that achieved classification accuracy of 76.30\% for two qubit framework (one victim)and 53.37\% for three qubit framework (two victims). The details and performance of the classifiers for additional frameworks are presented in section V. All the circuits were executed on VTT Q50 QC co-developed by VTT and IQM. 
The key contributions of this paper are:
\begin{enumerate}
    \item Demonstrate how crosstalk between qubit measurements can be used to leak information to adjacent qubits.
    \item Extend the study of measurement based crosstalk to different QC architecture. Although VTT Q50 is a superconducting quantum computer, its underlying hardware implementation is different from IBM processors, which was used by researchers in \cite{saki2021qubit}.
    \item Construct novel dataset based on the experiment data. The dataset contains the probability of measuring attacker qubit for different measurement scenarios. 
    \item Train and test SVM classifiers on the experimental datasets. The models have classification accuracy of 76.30\% for two qubit framework and 53.37\% for three qubit framework.
    
\end{enumerate}

The rest of the paper is organized as follows: Section II provides background information related to quantum computing. Section III reviews related work and Section IV describes the methodology including the attack framework used in our experiments. Section V presents the results of the experiment. Section VI discusses the results and opportunities for future research. Finally, Section VII concludes the paper.

\section{Background}
    \label{background}
    Quantum Computing utilizes core concepts of quantum mechanics like superposition and entanglement to solve complex operations that are challenging for classical computers. Superposition is the capability of a qubit to represent multiple states simultaneously. Whereas, entanglement is the correlation between qubits such that the state of one qubit affects the state of another.  

\subsection*{Qubits}
Qubit is the fundamental unit of quantum information. Contrary to the classical bit, it can exist in states 0 ($\ket{0}$), 1 ($\ket{1}$) and superposition of 0 and 1 ($\ket\Psi\ = \alpha\ket{0} +\beta\ket{1}$). The magnitudes of complex probability amplitudes $\alpha$ and $\beta$ determine the probabilities of the $\ket{0}$ and $\ket{1}$ states. A pure qubit state satisfies the normalization condition ($(|\alpha|^2)$ + $|\beta|^2$ =1) and can be visualized using a Bloch sphere. The Bloch sphere is a geometrical representation where a pure state corresponds to a point on the surface of the sphere and rotations about the axes represent operations on a qubit.

\subsection*{Quantum Gates}
Quantum gates are operations that can be performed on qubits. They are used to manipulate the qubit state and are analogous to logic gates in classical computing. Quantum gates are unitary and they preserve the normalization of a qubit state during transformation. The most common single qubit gates are Pauli X, Y, Z and Hadamard (H) whereas the most common two qubit gate is CNOT (CX). Most QC hardware have a few gates they can implement directly and are called native gates \cite{heng2022decomposition}. Other high-level gates are decomposed into native gates during transpilation.

\subsubsection*{Pauli X gate}
Pauli X is an operation that reverses the state of a qubit. 
The matrix representation of X gate is
\begin{equation*}
X = \begin{bmatrix}
0 & 1 \\1 & 0 \end{bmatrix}
\end{equation*}
\begin{equation*}
X\ket{0} =\ket{1} \ and\ X\ket{1} =\ket{0}
\end{equation*}

\subsubsection*{Pauli Y gate}
Pauli Y reverses the state of a qubit and introduces phase $i$. 
The matrix representation of Y gate is
\begin{equation*}
Y = \begin{bmatrix}
0 & -i \\i & 0 \end{bmatrix}
\end{equation*}
\begin{equation*}
Y\ket{0} =i\ket{1} \ and\ Y\ket{1} =-i\ket{0}
\end{equation*}

\subsubsection*{Pauli Z gate}
Pauli Z introduces phase $\pi$ between the basis states ($\ket{0}$ and $\ket{1}$).
The matrix representation of Z gate is
\begin{equation*}
Z = \begin{bmatrix}
1 & 0 \\0 & -1 \end{bmatrix}
\end{equation*}
\begin{equation*}
Z\ket{0} =\ket{0} \ and\ Z\ket{1} =-\ket{1}
\end{equation*}

\subsubsection*{Hadamard gate}
Hadamard operation transforms a qubit into equal superposition of the basis states.
The matrix representation of hadamard gate is
\begin{equation*}
H = 1/\sqrt{2}\begin{bmatrix}
1 & 1 \\1 & -1 \end{bmatrix}
\end{equation*}
\begin{equation*}
H\ket{0} =(\ket{0}+\ket{1})/\sqrt{2} \ and\ H\ket{1} =(\ket{0}-\ket{1})/\sqrt{2})
\end{equation*}

\subsubsection*{CNOT gate}
CNOT is a two qubit operation that reverses the target qubit based on the state of the control qubit. If the control qubit is $\ket{0}$, no operation is performed on the target qubit. If the control qubit is $\ket{1}$, the state of the target qubit is reversed.The matrix representation of CNOT gate is
\begin{equation*}
CX = \begin{bmatrix}
1 & 0 & 0 & 0 \\0 & 1 & 0 & 0\\0 & 0 & 0 & 1\\0 & 0 & 1 & 0\end{bmatrix}
\end{equation*}
\begin{equation*}
\begin{split}
CX\ket{00} =\ket{00} \ ,\  CX\ket{01} =\ket{01} \\ CX\ket{10} = \ket{11}\ ,\  CX\ket{11} = \ket{10}
\end{split}
\end{equation*}
where the first qubit is the control qubit and second qubit is the target.

\subsection*{Measurement}
Measurement or readout is the process of extracting classical information about a qubit (0 or 1) from its quantum state $\ket\Psi$. The outcome of the measurement is determined by the probability amplitudes ($\alpha$ and $\beta$ ). Measurement destroys the quantum information stored in a qubit and unlike quantum gates, qubit measurement is not reversible. To measure a qubit, it is coupled to a measuring device like a resonator or a photodetector \cite{barberena2024overview}.

\subsection*{CIA for Quantum Computers}
The CIA triad (Confidentialy, Integrity and Availibility) is the backbone of classical cybersecurity. We can extend CIA model to security of QCs with following considerations \cite{Adhikari2025review}:
\begin{enumerate}
    \item Confidentiality: An attacker should not be able to get any information about an algorithm that is being executed by another user in the same system.
    \item Integrity: An attacker should not be able to influence or degrade the fidelity of an algorithm that is being executed by another user in same the system.
    \item Availability: An attacker should not be able to trigger the QC system to go out of service for considerable amount of time.
\end{enumerate}

\section{Related Works}
    \label{literature} 
    Most works on security of QCs have utilized crosstalk to inject faults or leak information from the system \cite{ash2020analysis}\cite{choudhury2024}\cite{Lee2025}. A few others have used side channel leakage like power and timing traces to determine sensitive information about algorithms being executed \cite{xu2023exploration}\cite{erata2024quantum}\cite{lu2024quantum}. Operations on a qubit can interfere with adjacent qubits and the impact is especially significant with two qubit gates. The authors in \cite{ash2020analysis}\cite{deshpande2023design}\cite{ arellano2025qubitvise} demonstrated that executing multiple CNOT gates can decrease the fidelity of the output in adjacent qubits. Another work by \cite{mi2022securing} examined reset operation and found that residual information after reset can be used to predict the state of the qubit prior to reset.

The framework introduced by \cite{saki2021qubit} serves as the baseline for our work. Their study demonstrated that measuring a qubit in state `0' or `1' can have different impact on the measurement of the neighboring qubit. They used this framework to generate reference signatures that can be use to identify the state of neighboring qubit based on the measurement of a qubit. With the exception of a few studies, most experimental evaluations in this field, including \cite{saki2021qubit} have been conducted in IBM superconducting QCs. Our work expands on the methodology proposed in \cite{saki2021qubit} by adapting their framework to a different hardware platform (VTT Q50). This allows us to evaluate its validity as a potential security issue across a different QC hardware. We also extended the scope of the evaluation by introducing additional measurement scenarios, and integrating machine learning to classify the measured states.

\section{Methodology}
    \label{methodology} 
    \begin{figure}[htbp]
  \centering
  \captionsetup{font=footnotesize, width=\columnwidth} 
  \scalebox{0.60}{\begin{tikzpicture}[
    qubit/.style={
        circle,
        draw,
        thick,
        minimum size=8mm,
        inner sep=0pt,
        font=\normalsize
    }
]

\def\dx{1.0}
\def\dy{1.0}


\node[qubit] (q54) at (0*\dx,0*\dy) {q54};
\node[qubit] (q51) at (2*\dx,0*\dy) {q51};
\node[qubit] (q46) at (4*\dx,0*\dy) {q46};
\node[qubit] (q39) at (6*\dx,0*\dy) {q39};

\node[qubit] (q53) at (-1*\dx,-1*\dy) {q53};
\node[qubit] (q50) at (1*\dx,-1*\dy) {q50};
\node[qubit] (q45) at (3*\dx,-1*\dy) {q45};
\node[qubit] (q38) at (5*\dx,-1*\dy) {q38};
\node[qubit] (q31) at (7*\dx,-1*\dy) {q31};

\node[qubit] (q52) at (-2*\dx,-2*\dy) {q52};
\node[qubit] (q49) at (0*\dx,-2*\dy) {q49};
\node[qubit] (q44) at (2*\dx,-2*\dy) {q44};
\node[qubit] (q37) at (4*\dx,-2*\dy) {q37};
\node[qubit] (q30) at (6*\dx,-2*\dy) {q30};

\node[qubit] (q48) at (-1*\dx,-3*\dy) {q48};
\node[qubit] (q43) at (1*\dx,-3*\dy) {q43};
\node[qubit] (q36) at (3*\dx,-3*\dy) {q36};
\node[qubit] (q29) at (5*\dx,-3*\dy) {q29};
\node[qubit] (q22) at (7*\dx,-3*\dy) {q22};

\node[qubit] (q47) at (-2*\dx,-4*\dy) {q47};
\node[qubit] (q42) at (0*\dx,-4*\dy) {q42};
\node[qubit] (q35) at (2*\dx,-4*\dy) {q35};
\node[qubit] (q28) at (4*\dx,-4*\dy) {q28};
\node[qubit] (q21) at (6*\dx,-4*\dy) {q21};
\node[qubit] (q14) at (8*\dx,-4*\dy) {q14};

\node[qubit] (q41) at (-1*\dx,-5*\dy) {q41};
\node[qubit] (q34) at (1*\dx,-5*\dy) {q34};
\node[qubit] (q27) at (3*\dx,-5*\dy) {q27};
\node[qubit] (q20) at (5*\dx,-5*\dy) {q20};
\node[qubit] (q13) at (7*\dx,-5*\dy) {q13};

\node[qubit] (q40) at (-2*\dx,-6*\dy) {q40};
\node[qubit] (q33) at (0*\dx,-6*\dy) {q33};
\node[qubit] (q26) at (2*\dx,-6*\dy) {q26};
\node[qubit] (q19) at (4*\dx,-6*\dy) {q19};
\node[qubit] (q12) at (6*\dx,-6*\dy) {q12};
\node[qubit] (q7)  at (8*\dx,-6*\dy) {q7};

\node[qubit] (q32) at (-1*\dx,-7*\dy) {q32};
\node[qubit] (q25) at (1*\dx,-7*\dy) {q25};
\node[qubit] (q18) at (3*\dx,-7*\dy) {q18};
\node[qubit] (q11) at (5*\dx,-7*\dy) {q11};
\node[qubit] (q6)  at (7*\dx,-7*\dy) {q6};

\node[qubit] (q24) at (0*\dx,-8*\dy) {q24};
\node[qubit] (q17) at (2*\dx,-8*\dy) {q17};
\node[qubit] (q10) at (4*\dx,-8*\dy) {q10};
\node[qubit, text=blue] (q5)  at (6*\dx,-8*\dy) {q5};
\node[qubit] (q2)  at (8*\dx,-8*\dy) {q2};

\node[qubit] (q23) at (-1*\dx,-9*\dy) {q23};
\node[qubit] (q16) at (1*\dx,-9*\dy) {q16};
\node[qubit] (q9)  at (3*\dx,-9*\dy) {q9};
\node[qubit, text=red] (q4)  at (5*\dx,-9*\dy) {q4};
\node[qubit] (q1)  at (7*\dx,-9*\dy) {q1};

\node[qubit] (q15) at (0*\dx,-10*\dy) {q15};
\node[qubit] (q8)  at (2*\dx,-10*\dy) {q8};
\node[qubit, text=blue] (q3)  at (4*\dx,-10*\dy) {q3};


\foreach \a/\b in {
q54/q53,q54/q50,q51/q50,q51/q45,q46/q45,q46/q38,q39/q38,q39/q31,q53/q52,q53/q49,q50/q49,q50/q44,q45/q44,q45/q37,q38/q37,q38/q30,q31/q30,q52/q48,q49/q48,q49/q43,q44/q43,q44/q36,q37/q36,q37/q29,q30/q29,q30/q22,q48/q47,q48/q42,q43/q42,q43/q35,q36/q35,q36/q28,q29/q28,q29/q21,q22/q21,q22/q14,q47/q41,q42/q41,q42/q34,q35/q34,q35/q27,q28/q27,q28/q20,q21/q20,q21/q13,q14/q13,q41/q40,q41/q33,
q34/q33,q34/q26,q27/q26,q27/q19,q20/q19,q20/q12,q13/q12,q13/q7,q40/q32,q33/q32,q33/q25,q26/q25,q26/q18,q19/q18,q19/q11,q12/q11,q12/q6,q7/q6,q32/q24,q25/q24,q25/q17,q18/q17,q18/q10,q11/q10,q11/q5,q6/q5,q6/q2,q24/q23,q24/q16,q17/q16,q17/q9,q10/q9,q10/q4,q5/q4,q5/q1,q23/q15,q16/q15,q16/q8,q9/q8,q9/q3,q4/q3,q2/q1}
{
    \draw (\a) -- (\b);
}

\end{tikzpicture}}
  \caption{Architecture of VTT Q50 with physical mapping of victim and attacker qubits. Q0 corresponds to physical qubit q3, Q1 corresponds to q4 and Q2 corresponds to q5 in the quantum processor.}
  \label{fig:Architecture of VTTQ50}
\end{figure}
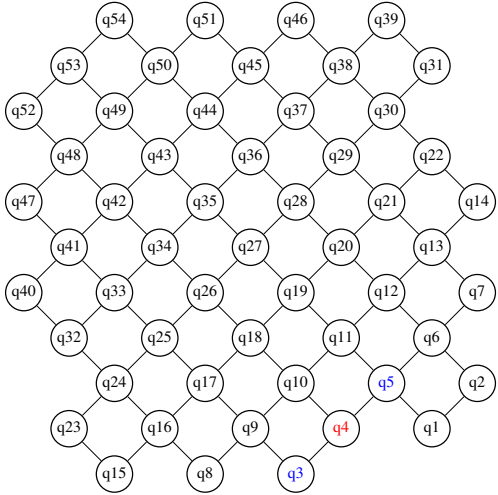

This work builds on the methodology introduced by the authors in \cite{saki2021qubit}. They studied information leak from crosstalk generated through measurement of neighboring qubit. For, two qubit system, one qubit is designated as the victim and the other is the attacker qubit. The qubit whose state is of interest for prediction is the victim qubit. The qubit that is affected by the variation in measurement of the victim qubit is the attacker qubit \cite{saki2021qubit}.
For our frameworks, the specifications of the victim and attacker qubits are:
\subsection*{Victim Qubits (VQ)}
Victim Qubits are measured in three different configurations `0', `1' and no measurement. For experiments with two qubits, the victim and attacker qubits (Q0 and Q1 respectively) are adjacent to each other. For three qubits (victim Q0, attacker Q1 and victim Q2), the attacker qubit is sandwiched between two victim qubits. The visual representation of the victim and attacker qubits in the quantum processor is depicted in  Figure~\ref{fig:Architecture of VTTQ50}.

\subsection*{Attacker Qubit (AQ)}
Ideally, the attacker qubit should always be measured `1'. However, in noisy quantum processors, the probability of measuring attacker qubit `1' changes due to noise from adjacent measurements. This difference in the measurement of the attacker qubit based on the measurement of victim qubit can be used to predict the victim qubit.\\

To evaluate measurement based crosstalk, we designed experiments across four different cases. 

\begin{table}[htbp]
\renewcommand{\arraystretch}{1.2} 
\captionsetup{font=footnotesize, justification=centering}
\caption{Measurement configurations for all analyzed frameworks}
\label{tab:all_scenarios}
\centering
\fontsize{6.5pt}{7.5pt}\selectfont
\begin{tabular}{cccc}
\hline
\bfseries Q0 measurement & \bfseries Q1 measurement & \bfseries Q2 measurement & \bfseries label \\
\hline
\multicolumn{4}{c}{Two Qubit Framework} \\\hline
0 & 1 & - & 0  \\
1 & 1 & - & 1\\
No measurement & 1 & - & no \\ \hline
\multicolumn{4}{c}{Three Qubit Framework} \\\hline
0 & 1 & 0 & 0\_0 \\
0 & 1 & 1 & 0\_1 \\
1 & 1 & 1 & 1\_1 \\
No measurement & 1 & No measurement & no\_no \\
No measurement & 1 & 0 & no\_0 \\
No measurement & 1 & 1 & no\_1 \\ \hline
\multicolumn{4}{c}{Measurement Count Framework} \\\hline
0 & 1 & 0 & \multirow{3}{*}{Two VQ measurement} \\
0 & 1 & 1 \\
1 & 1 & 1 \\ \hline
No measurement & 1 & 1 &\multirow{2}{*}{One VQ measurement} \\
No measurement & 1 & 0 \\ \hline
No measurement & 1 & No measurement  & No VQ Measurement\\
\hline
\end{tabular}
\end{table}

\subsection*{Case 1: Two Qubit Framework}
For two qubit framework, we examined three scenarios. These include measuring the victim qubit `0', `1' and no measurement. The measurement scenarios with corresponding labels for two qubit framework is presented in Table~\ref{tab:all_scenarios}. In this framework, we collected data from execution of 900 circuits. All the circuits were executed within the same calibration window of Q50 and each circuit was executed for 1000 shots. The circuits were composed of 1-10 X gates on the victim qubit and single X gate on the attacker qubit.

\subsection*{Case 2: Two Qubit Framework with mixed calibration data}
The primary distinction between Case 1 and Case 2 is that in this case, data collection is not limited to single calibration window of Q50. We collected the results from 900 circuits executed during two calibration windows (450 for each calibration window). Each circuit was executed for 1000 shots. The derived dataset uses the same scenarios and labels as Case 1 presented in Table~\ref{tab:all_scenarios}.

\subsection*{Case 3: Three Qubit Framework}
In three qubit framework, there are two victim qubits and single attacker qubit between them. We examined six configurations for this case. The scenarios are listed in Table~\ref{tab:all_scenarios}. We collected data from 1800 circuits, with each configuration corresponding to 300 observations. The circuits were composed of 1-10 X gates on the victim qubits and single X gate on the attacker qubit.

\subsection*{Case 4: Measurement Count Framework}
The measurement count framework evaluates the correlation between the number of qubits measured and measurement of adjacent qubit. The difference between Case 4 and all the previous cases is that the measurement outcome `0' or `1' is not considered a classification category. Instead the observations are classified based on the number of neighboring qubits measured (1 or 2). The scenarios for this framework are listed in Table~\ref{tab:all_scenarios}.

\subsection*{Evaluation Platform}
\begin{figure*}[t]
\captionsetup{font=footnotesize} 
     \begin{subfigure}[c]{0.30\textwidth}
     \begin{subfigure}[c]{\textwidth}
         \resizebox{\textwidth}{!}{%
         \begin{tikzcd}[sep=small]
            \lstick{$\ket{q_0}$} & \gate{x} & \gate{x} & \gate{x} & \gate{x} & \qw & \qw & \meter{} \\
            \lstick{$\ket{q_1}$} & \gate{x} & \qw & \qw & \qw & \qw & \qw & \meter{}
         \end{tikzcd}
         }
        \caption{}
         \label{fig:circuit 00}
     \end{subfigure}
    \\ 
     \begin{subfigure}[c]{\textwidth}
            \resizebox{\textwidth}{!}{%
            \begin{tikzcd}[sep=small]
            \lstick{$\ket{q_0}$} & \gate{x} & \gate{x} & \gate{x} & \gate{x} & \gate{x} & \meter{} \\
            \lstick{$\ket{q_1}$} & \gate{x}  & \qw & \qw & \qw & \qw & \meter{}
         \end{tikzcd}
         }
         \caption{}
         \label{fig:circuit 11}
     \end{subfigure}
     \\
     \begin{subfigure}[c]{\textwidth}
            \resizebox{\textwidth}{!}{%
            \begin{tikzcd}[sep=small]
            \lstick{$\ket{q_0}$} & \qw & \qw & \qw & \qw & \qw & \qw & \qw & \qw & \qw & \qw & \qw \\
            \lstick{$\ket{q_1}$} & \gate{x}  & \qw & \qw & \qw & \qw & \qw & \qw & \qw & \qw & \qw & \meter{}
         \end{tikzcd}
         }
         \caption{}
         \label{fig:circuit 10}
     \end{subfigure}
     \end{subfigure}
     \hfill
    \begin{subfigure}[c]{0.30\textwidth}    
     \begin{subfigure}[c]{\textwidth}
            \resizebox{\textwidth}{!}{%
            \begin{tikzcd}[sep=small]
             \lstick{$\ket{q_0}$} & \gate{x} & \gate{x} & \gate{x} & \gate{x} & \qw & \qw & \meter{} \\
            \lstick{$\ket{q_1}$} & \gate{x}  & \qw & \qw & \qw &   \qw & \qw & \meter{}\\
             \lstick{$\ket{q_2}$} & \gate{x} & \gate{x} & \gate{x} & \gate{x} & \qw & \qw & \meter{} \\
         \end{tikzcd}
         }
         \caption{}
         \label{fig:circuit 10}
     \end{subfigure}
     \\
     \begin{subfigure}[c]{\textwidth}
       
            \resizebox{\textwidth}{!}{%
            \begin{tikzcd}[sep=small]
             \lstick{$\ket{q_0}$} & \gate{x} & \gate{x} & \gate{x} & \gate{x} & \qw & \meter{} \\
            \lstick{$\ket{q_1}$} & \gate{x}  & \qw & \qw & \qw & \qw & \meter{}\\
            \lstick{$\ket{q_2}$} & \gate{x} & \gate{x} & \gate{x} & \gate{x} & \gate{x} & \meter{}
         \end{tikzcd}
         }
         \caption{}
         \label{fig:circuit 10}
     \end{subfigure}
\\
     \begin{subfigure}[c]{\textwidth}
            \resizebox{\textwidth}{!}{%
            \begin{tikzcd}[sep=small]
            \lstick{$\ket{q_0}$} & \gate{x} & \gate{x} & \gate{x} & \gate{x} & \gate{x} & \meter{} \\
            \lstick{$\ket{q_1}$} & \gate{x}  & \qw & \qw & \qw & \qw & \meter{}\\
            \lstick{$\ket{q_2}$} & \gate{x} & \gate{x} & \gate{x} & \gate{x} & \gate{x} & \meter{}
         \end{tikzcd}
         }
         \caption{}
         \label{fig:circuit 10}
     \end{subfigure}
     \end{subfigure}
  \hfill
     \begin{subfigure}[c]{0.30\textwidth}
     \begin{subfigure}[c]{\textwidth}
        
            \resizebox{\textwidth}{!}{%
            \begin{tikzcd}[sep=small]
            \lstick{$\ket{q_0}$} & \qw  & \qw & \qw & \qw & \qw & \qw & \qw & \qw & \qw & \qw & \qw & \\
            \lstick{$\ket{q_1}$} & \gate{x}  & \qw & \qw & \qw & \qw & \qw & \qw & \qw & \qw & \qw & \meter{}\\
            \lstick{$\ket{q_2}$} & \qw  & \qw & \qw & \qw & \qw & \qw & \qw & \qw & \qw & \qw & \qw & \\
         \end{tikzcd}
         }
         \caption{}
         \label{fig:circuit 10}
     \end{subfigure}
     \\
     \begin{subfigure}[c]{\textwidth}
        
            \resizebox{\textwidth}{!}{%
            \begin{tikzcd}[sep=small]
            \lstick{$\ket{q_0}$} & \qw  & \qw & \qw & \qw & \qw & \qw & \qw &  \\
            \lstick{$\ket{q_1}$} & \gate{x}  & \qw & \qw & \qw &  \qw & \qw & \meter{}\\
            \lstick{$\ket{q_2}$} & \gate{x} & \gate{x} & \gate{x} & \gate{x} & \qw & \qw & \meter{}
         \end{tikzcd}
         }
         \caption{}
         \label{fig:circuit 10}
     \end{subfigure}
     \\
     \begin{subfigure}[c]{\textwidth}
        
            \resizebox{\textwidth}{!}{%
            \begin{tikzcd}[sep=small]
            \lstick{$\ket{q_0}$} & \qw  & \qw & \qw & \qw & \qw & \qw & \qw  \\
            \lstick{$\ket{q_1}$} & \gate{x}  & \qw & \qw & \qw &  \qw & \qw & \meter{}\\
            \lstick{$\ket{q_2}$} & \gate{x} & \gate{x} & \gate{x} & \gate{x} & \gate{x} & \qw & \meter{}
         \end{tikzcd}
         }
         \caption{}
         \label{fig:circuit 10}
     \end{subfigure}
     \end{subfigure}
     \\
     \caption{Representative circuits for two qubit frameworks (a-c), three qubit framework and measurement count framework (d-i) . The victim qubit is subjected to 1-10 X gates. The circuits consist of odd number of X gates for victim measurement `1' and even number of X gates for victim measurement `0'.  For no measurement, no operation is performed and the victim is not measured. (a) Circuit for measuring victim qubit `0' and attacker qubit `1'. (b)  Circuit for measuring victim qubit`1' and attacker qubit `1'. (c) Circuit for no measurement on victim qubit and measuring attacker qubit `1'. (d) Circuit for measuring both victim qubits `0' and attacker qubit `1'. (e) Circuit for measuring one victim qubit `0', another victim qubit `1' and attacker qubit `1'. (f) Circuit for measuring both victim qubits `1' and attacker qubit `1'. (g) Circuit for no measurement on either victim qubits and measuring attacker qubit `1'. (h) Circuit for no measurement on one victim qubit, measuring another victim qubit `0' and attacker qubit `1'. (i) Circuit for no measurement on one victim qubit, measuring another victim qubit `0' and attacker qubit `1'.}
     \label{fig:all_circuits}
\end{figure*}
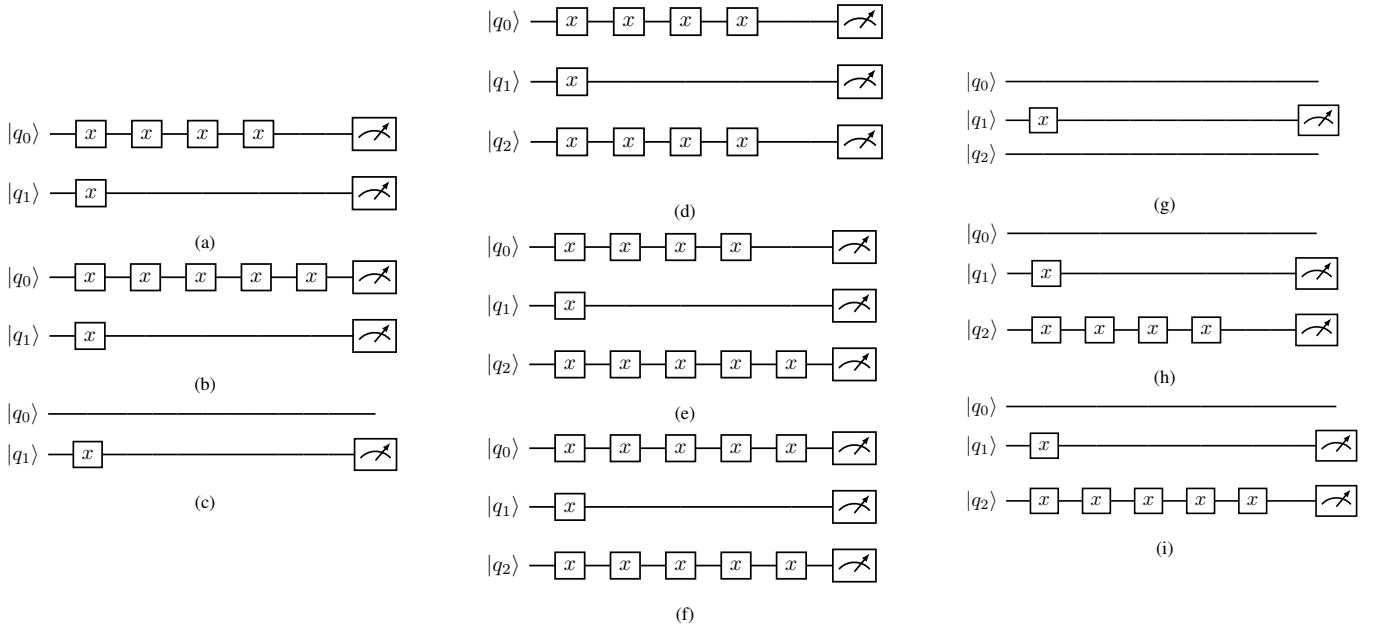

All the circuits were executed in VTT Q50 QC. Q50 is a superconducting QC with 53 qubits arranged in a lattice topology. The qubit layout of Q50 is illustrated in Figure~\ref{fig:Architecture of VTTQ50}. The representative circuits for different measurement cases are presented in Figure~\ref{fig:all_circuits}.

\subsection*{Data Collection}
Data collection for this work was performed by executing quantum circuits in superconducting quantum hardware. The circuits for all the cases except Case 2 were executed consecutively within a single calibration window to minimize interference from calibration.

The outputs from circuit executions were used to construct experimental datasets in all the cases. The datasets were used to train and test SVM models. The results from classification models are presented in Section V.

\section{Experimental Results}
 \label{results} 
    
\begin{figure*}[htbp]
\captionsetup{font=footnotesize, justification=centering} 
\begin{subfigure}[c]{0.45\textwidth}
  \centering
  \captionsetup{font=scriptsize, justification=justified} 
   \includegraphics[width=\textwidth]{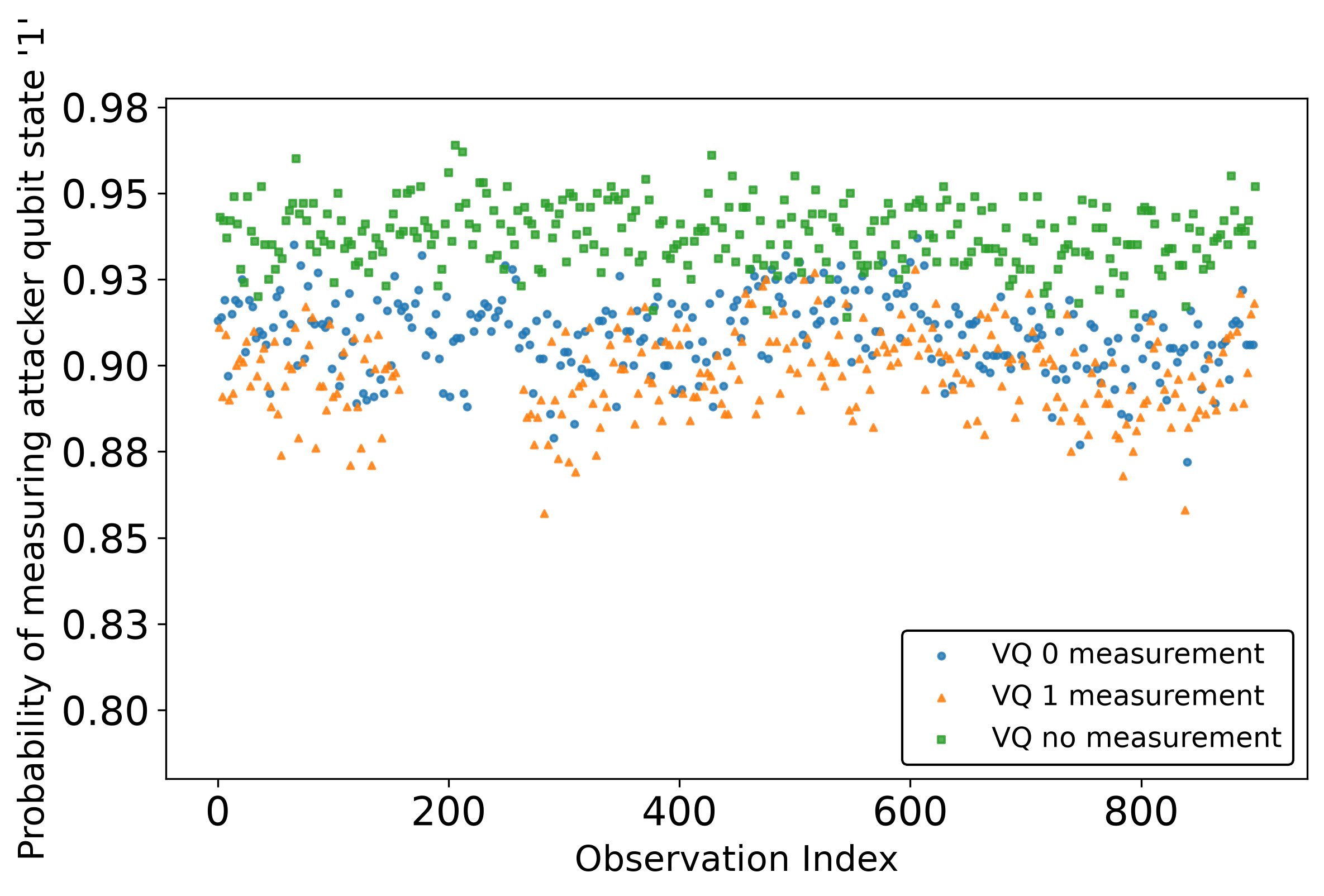}
  \caption{Probability of measuring attacker qubit `1' for independent experiments in two qubit framework}
  \label{fig: scatter_plot_two_qubits}
\end{subfigure}
\hfill
\begin{subfigure}[c]{0.45\textwidth}
  \centering
  \captionsetup{font=scriptsize, justification=justified}
  \includegraphics[width=\textwidth]{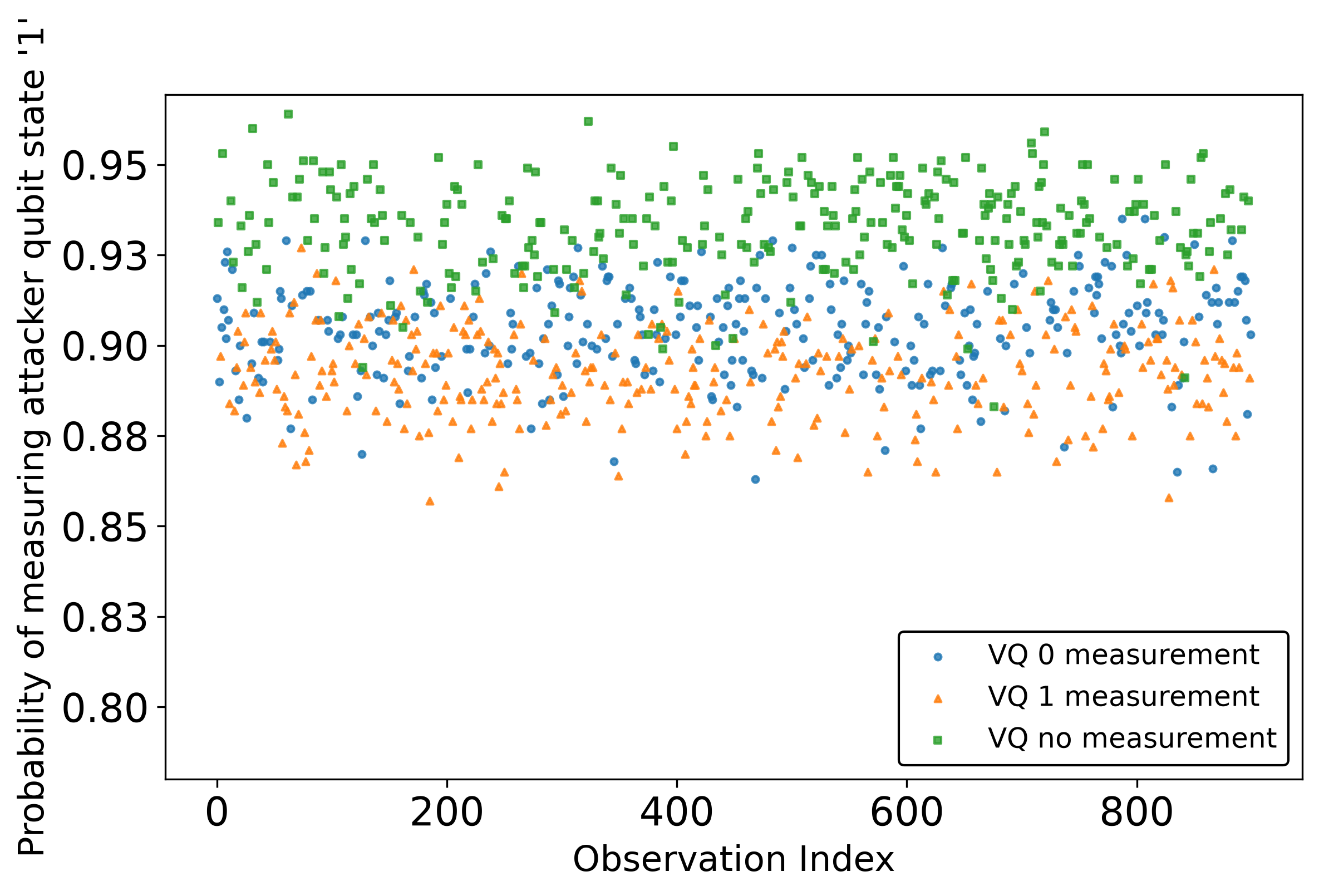}
  \caption{Probability of measuring attacker qubit `1' for independent experiments in mixed calibration framework}
  \label{fig: scatter_plot_two_qubits_cross_calibration}
\end{subfigure}
\\
\begin{subfigure}[c]{0.45\textwidth}
  \centering
  \captionsetup{font=scriptsize, justification=justified} 
   \includegraphics[width=\textwidth]{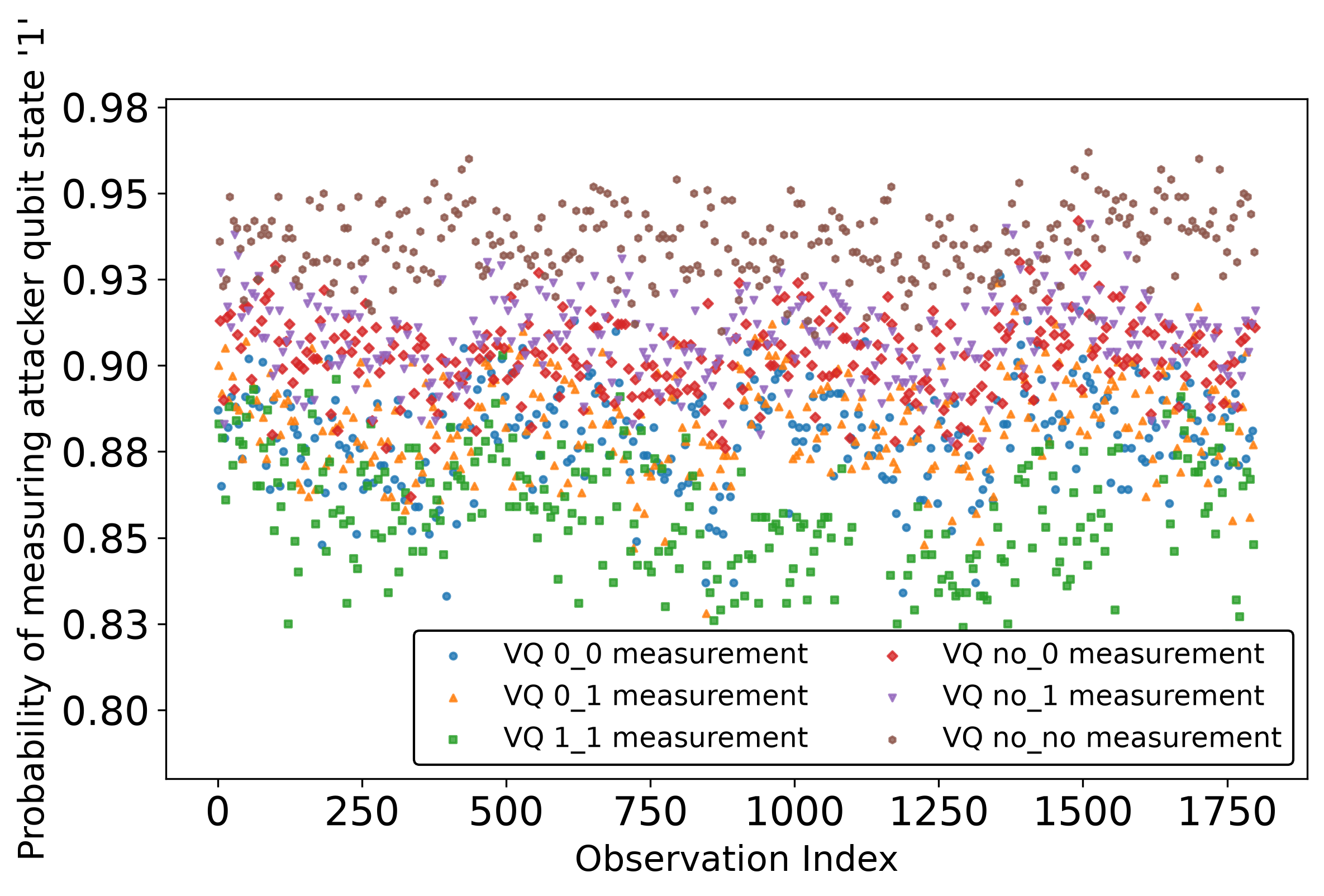}
  \caption{Probability of measuring attacker qubit `1' for independent experiments in three qubit framework}
  \label{fig: scatter_plot_three_qubits}
\end{subfigure}
\hfill
\begin{subfigure}[c]{0.45\textwidth}
  \centering
  \captionsetup{font=scriptsize, justification=justified} 
   \includegraphics[width=\textwidth]{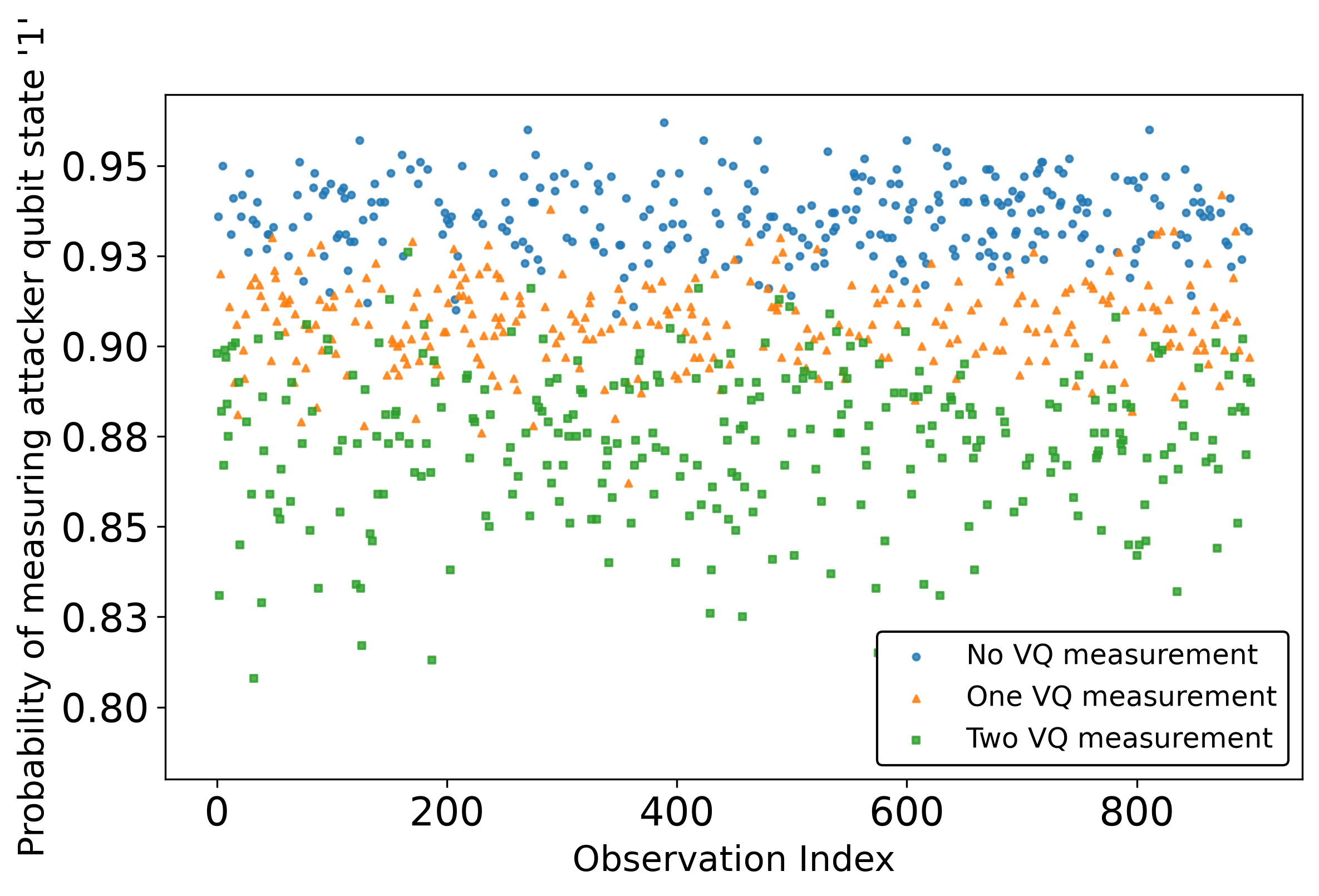}
  \caption{Probability of measuring attacker qubit `1' for independent experiments in measurement count framework}
  \label{fig: scatter_plot_measurement_count}
\end{subfigure}
\caption{Scatter plots illustrating the probability of measuring attacker qubit `1' for independent experiments in all analyzed frameworks}
\label{fig: scatter_plot_all}
\end{figure*}

This section provides the results from the experiments and the performance metrics of the machine learning models. 

\subsection*{Case 1: Two Qubit Framework}
For case 1, we collected execution data from 900 independent circuits. The circuit suite consisted of 300 circuits from each scenario listed in Table~\ref{tab:all_scenarios}. Each circuit was executed for 1000 shots and the probability of measuring the attacker qubit `1' was recorded. The probability was designated as an input feature in the dataset while victim measurement served as the classification label. The initial dataset consisted of 900 observations. Preliminary data auditing indicated that 35 observations deviated significantly from the primary data distribution. These observations were all from the scenario measuring victim qubit `1'. These observations were excluded from the final dataset to prevent boundary distortion and overfitting. The resulting final data set consisted of 865 observations. The dataset was stratified and split into training (80\%) and test (20\%) datasets. Classification was performed utilizing SVM equipped with a Radial Basis Function (RBF) kernel. \\

\begin{table}[htbp]
\renewcommand{\arraystretch}{1.2} 
\captionsetup{font=footnotesize, justification=centering}
\caption{Performance metrics for all analyzed frameworks}
\label{tab:all performance metrics}
\centering
\begin{tabular}{ccccc}
\hline
\bfseries Framework &\bfseries Accuracy & \bfseries Precision & \bfseries Recall & \bfseries F1-score \\
\hline
Two qubit & 0.7630 & 0.7724 & 0.7553 & 0.7570\\
Mixed calibration &0.7389 &  0.7472 & 0.7389 & 0.7422 \\
Three qubit & 0.5337 & 0.5358 & 0.5347 & 0.5324 \\
Measurement Count & 0.8611 & 0.8719 & 0.8611 & 0.8624 \\
\hline
\end{tabular}
\end{table}

\subsubsection*{Results}
The distribution of measuring attacker qubit `1' across independent experiments is depicted in Figure~\ref{fig: scatter_plot_two_qubits}. A discernible separation is observable between the data points that correspond to different victim measurements (`0', `1' and no measurement). The optimized SVM model trained on the final dataset achieved an accuracy of 76.30\% on classifying the victim measurement. The other performance metrics of the model are presented in Table~\ref{tab:all performance metrics}. The classification accuracy of the model on test dataset is displayed in Figure~\ref{fig: confusion matrix for two qubit classifier}.

\subsection*{Case 2 : Two Qubit Framework With Mixed Calibration Data}
In this framework, mixed calibration dataset was used to train and test the classifier. The mixed dataset contained results from circuit executions during two different calibration windows of Q50. The dataset included a total of 900 observations, 450 observations belonging to each calibration window. Out of the 900 circuits, there were 300 circuits from each of the three scenarios listed in Table~\ref{tab:all_scenarios}. The dataset was divided into training and test datasets with 80/20 split and used to train and test RBF SVM classifier. 

\subsubsection*{Results}
\begin{figure*}[htbp]
\captionsetup{font=footnotesize, justification=centering} 
\begin{subfigure}[c]{0.24\textwidth}
  \centering
  \captionsetup{font=scriptsize, justification=justified} 
   \includegraphics[width=\textwidth]{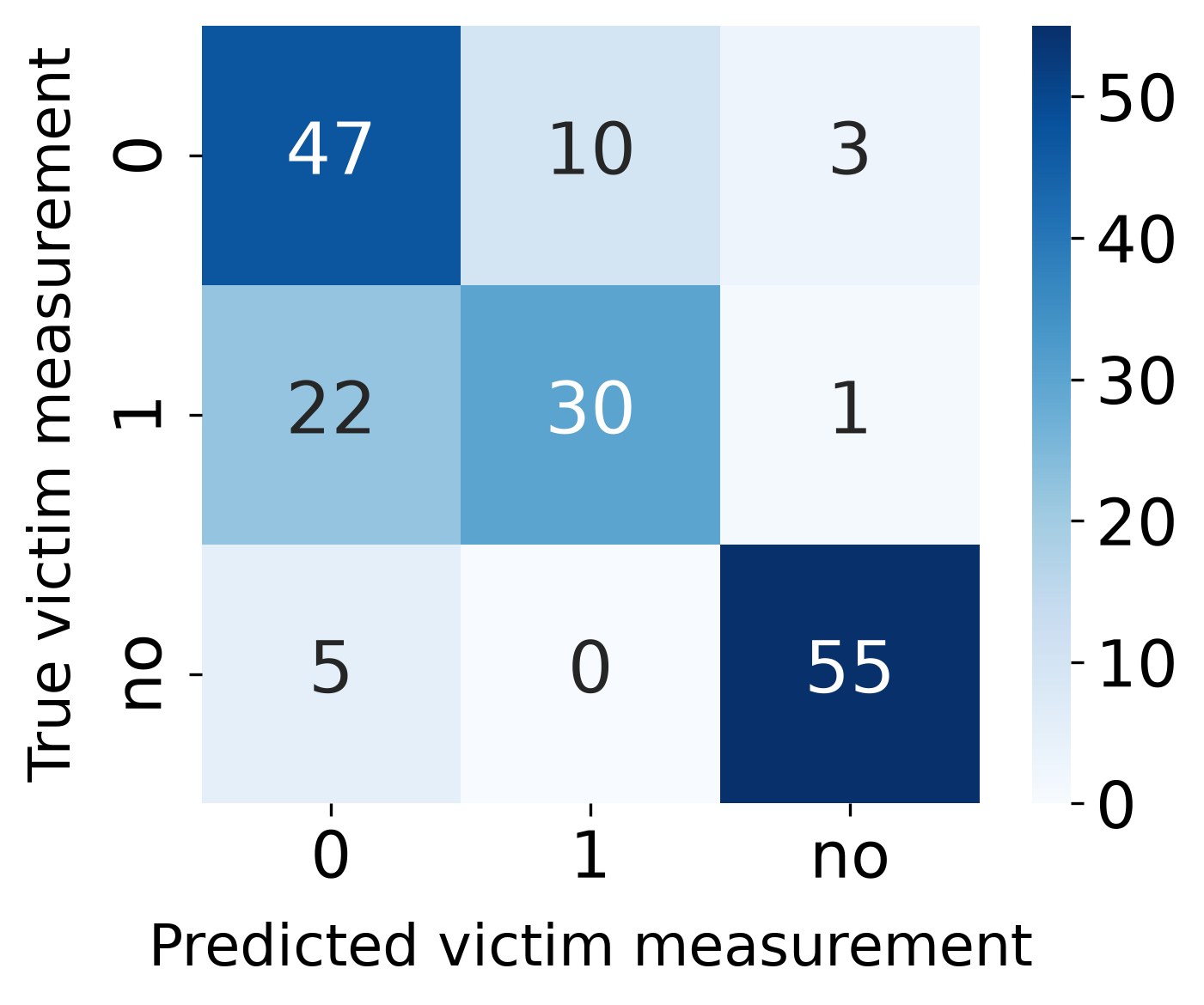}
  \caption{Confusion matrix evaluated on the test dataset for two qubit framework}
  \label{fig: confusion matrix for two qubit classifier}
\end{subfigure}
\hfill
\begin{subfigure}[c]{0.24\textwidth}
  \centering
  \captionsetup{font=scriptsize, justification=justified} 
   \includegraphics[width=\textwidth]{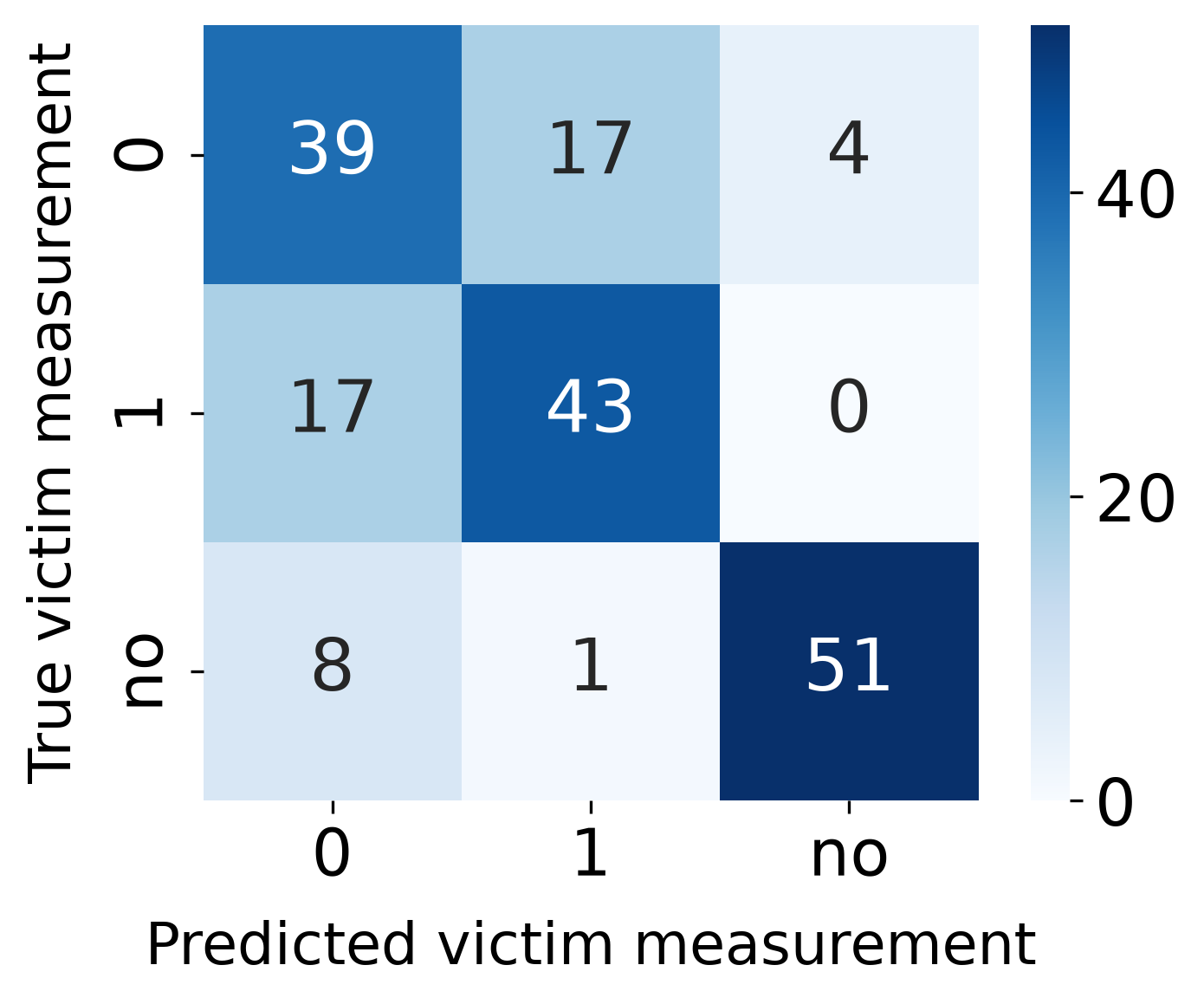}
  \caption{Confusion matrix evaluated on test dataset for mixed calibration framework} 
  \label{fig: confusion matrix of mixed collaboration data}
\end{subfigure}
\hfill
\begin{subfigure}[c]{0.24\textwidth}
  \centering
  \captionsetup{font=scriptsize, justification=justified} 
   \includegraphics[width=\textwidth]{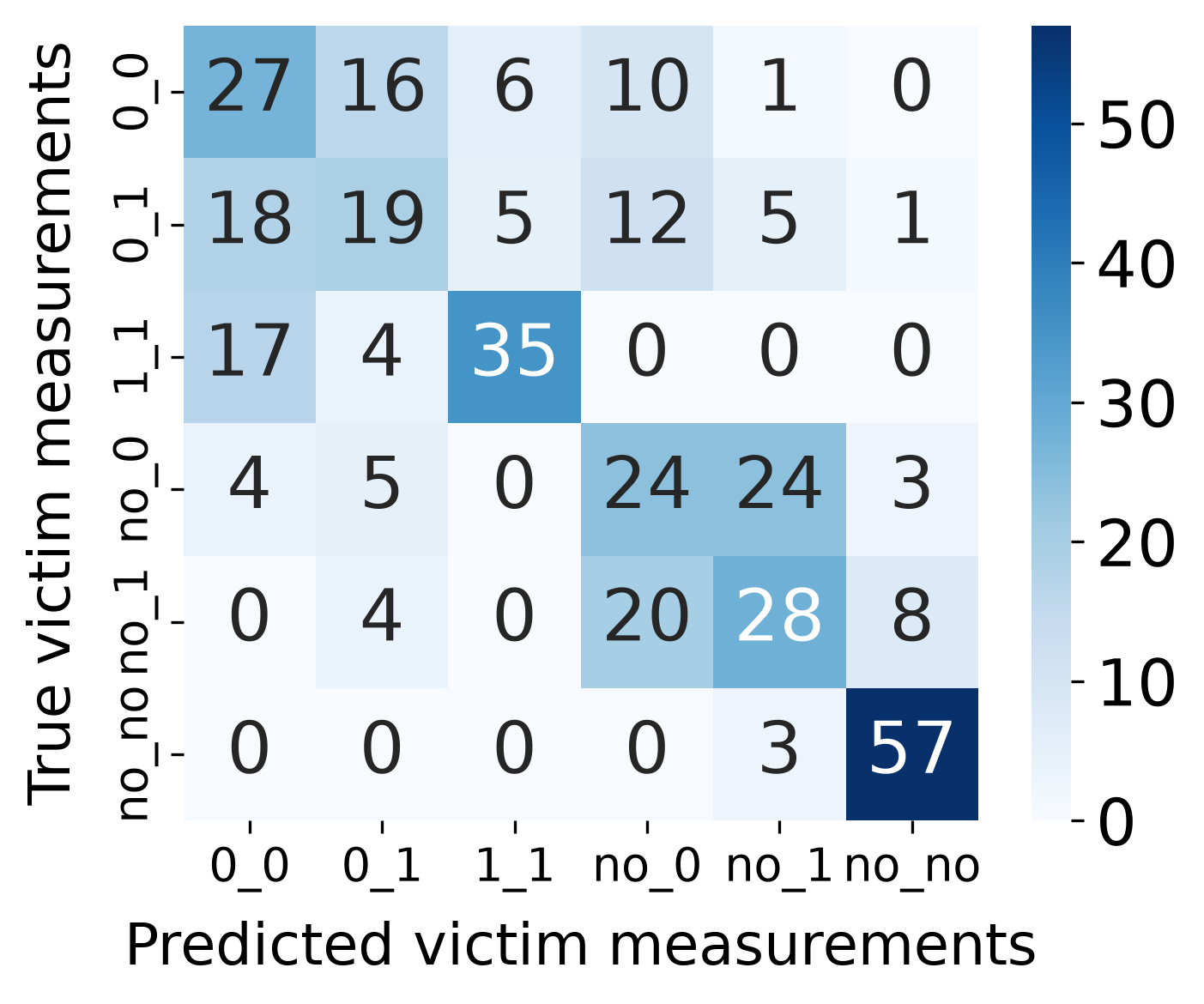}
  \caption{Confusion matrix evaluated on the test dataset for three qubit framework.}
  \label{fig: confusion matrix of three qubit measurement}
\end{subfigure}
\hfill
\begin{subfigure}[c]{0.24\textwidth}
  \centering
  \captionsetup{font=scriptsize, justification=justified} 
   \includegraphics[width=\textwidth]{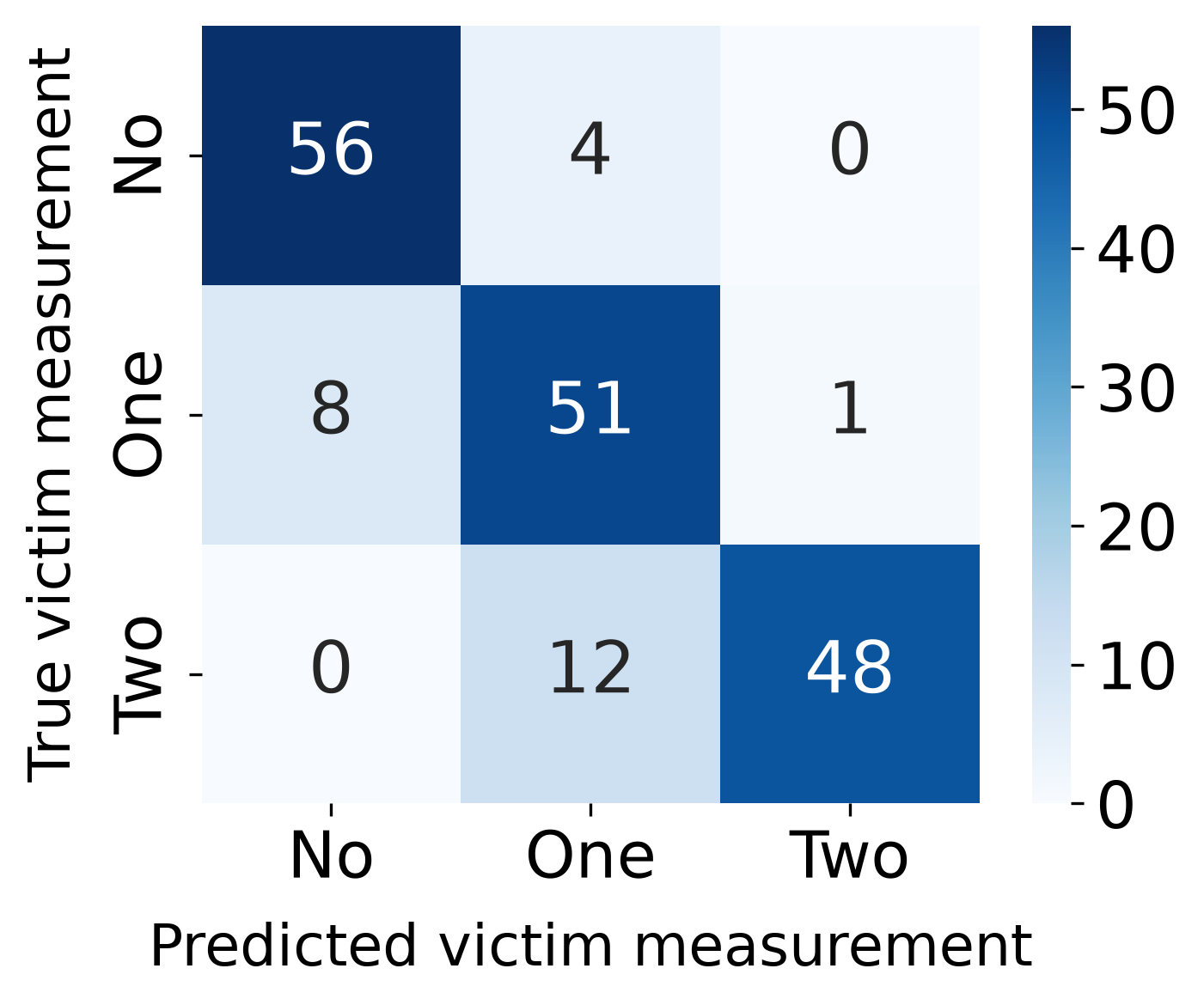}
  \caption{Confusion matrix evaluated on the test dataset for measurement count framework} 
  \label{fig: confusion matrix of modified three qubit measurement}
\end{subfigure}
  \caption{Confusion matrix evaluated on the test dataset for all frameworks} 
  \label{fig: confusion matrix}
\end{figure*}

The scatter plot in Figure~\ref{fig: scatter_plot_two_qubits_cross_calibration} illustrates the probability of measuring attacker qubit `1' for different victim measurements. The observations of the three classes exhibit moderate overlap, though distinct clusters remain visible. The SVM classifier trained on this data achieved an accuracy of 73.89\% on classifying the victim measurement. The other performance metrics are presented in Table~\ref{tab:all performance metrics}. The corresponding confusion matrix for classifying the test dataset is depicted in Figure~\ref{fig: confusion matrix of mixed collaboration data}.

\subsection*{Case 3: Three Qubit Framework}
In three qubit framework, we analyzed the impact of measuring two victim qubits on one attacker qubit. For this, we collected execution data from 1800 circuits. The total circuits included 300 circuits of each scenario presented in Table~\ref{tab:all_scenarios}. Similar to two qubit framework, each circuit was executed for 1000 shots and the probability of measuring the attacker qubit `1' was extracted. The generated dataset utilized the probability as a characteristic feature and the victim measurement scenarios as classification targets.  The dataset was partitioned into training and test datasets with an 80/20 split and was used to train and test a RBF SVM model.

The initial dataset consisted of 1800 observations derived from the execution of the circuits. We filtered the dataset to exclude 22 anomalous data points and the final dataset consisted of 1778 observations. The observations that were excluded belonged to the class that measures both victim qubits `1' (1\_1). The final dataset was partitioned into training and test datasets with an 80/20 split and was used to train and test a RBF SVM model.

\subsubsection*{Results}
The accuracy of optimized SVM classifier trained on three qubit dataset is 53.37\%. Other performance metrics of the classifier are presented in Table~\ref{tab:all performance metrics}. Compared to Case 1 and Case 2, there is a noticeable decline in the accuracy of the classifier for two victim qubits.   The drop in accuracy is also evident from the data distribution in Figure~\ref{fig: scatter_plot_three_qubits}. The plot shows distribution measuring attacker qubit `1' (probability) for different victim measurement classes. In the graph, there is a high degree of overlap in the data points for some victim measurement categories. For example, the class where both victims are measured `0' and the class where one victim is measured `0' and another is measured `1' (0\_0 and 0\_1) exhibit poor separation. The confusion matrix depicted in Figure~\ref{fig: confusion matrix of three qubit measurement} shows that the classifier achieved high accuracy in classifying instances where neither victim qubit was measured (no\_no) and where both victim qubits were measured `1' (1\_1).
 
\subsection*{Case 4 : Measurement Count Framework} 
Although there is a high degree of overlap between data points of some classes (eg., 0\_0 and 1\_1) in Figure~\ref{fig: scatter_plot_three_qubits}, others maintain distinct boundaries with limited intersection (eg., no\_no and 0\_0). To further analyze this, we modified the dataset used in the three qubit framework. The modified dataset comprises of three class labels based on the number of victim qubits measured - one victim qubit measured, two victim qubits measured and no victim qubits measured. The classification only considers the number of qubits measured and remains invariant to the state of the qubits measured (`0' or `1'). The modification involved combining observations of the original dataset in Case 3 to fit into new classes. The regrouping of the observations with new labels is detailed in Table~\ref{tab:all performance metrics}. So, the final dataset contained 900 observations sampled from dataset of Case 3 and three labels. The final dataset was split into training and test datasets with an 80/20 ratio and was used to train and test a RBF SVM classifier.

\subsubsection*{Results}
The scatter plot in Figure~\ref{fig: scatter_plot_measurement_count} shows the data distribution with new classes. It is clear from the plot that there is a separation between the data in the three classes with limited overlap. The trained SVM classifier achieves an accuracy of 86.11\% on classifying the number of victim measurements. The other performance metrics of the classifier are listed in Table~\ref{tab:all performance metrics}. The confusion matrix for the classification of test dataset is illustrated in Figure~\ref{fig: confusion matrix of modified three qubit measurement}.

\subsection*{Classification using Jensen-Shannon Distance (JSD)} 
In addition to SVM, we utilized JSD to create a comparative baseline to the previous literature \cite{saki2021qubit}. This entailed building reference signatures for the observations measuring single victim qubit `0' and `1'. The dataset does not include observations where no measurement operation is performed on the victim qubit. From the sample of 565 observations, 452 observations were used for generating reference signatures and the remaining 113 observations were reserved for testing. We achieved an accuracy of 71.68\% in classifying the measurement of one victim qubit with JSD.

\section{Discussion}
 \label{discussion}
    The scatter plots in Figure~\ref{fig: scatter_plot_two_qubits}, \ref{fig: scatter_plot_two_qubits_cross_calibration} and \ref{fig: scatter_plot_measurement_count} exhibit a discernible degree of separation across observations of different classes. The segregation is more pronounced in Figure~\ref{fig: scatter_plot_measurement_count}, where the classes correspond to only the number of qubits measured irrespective of the state (`0' or `1'). Here, The data distribution illustrates a hierarchy with measurement of both victim qubits exhibiting the highest readout error on the attacker qubit, followed by measurement of one qubit and no measurement of either qubit. For frameworks with one victim qubit (Figure~\ref{fig: scatter_plot_two_qubits} and \ref{fig: scatter_plot_two_qubits_cross_calibration}), measuring `1' on the victim qubit generally corresponds to a higher readout error on the attacker qubit although there is some overlap between the observations measuring victim qubit `0' and `1'. However, increasing the number of victim qubits to two, creates significant overlap for some classes whereas others cluster with limited overlap. In all cases, not measuring the victim qubits yield the lowest readout error on the attacker qubit.

The authors in \cite{saki2021qubit} conducted their experiments in IBM QCs and reported an accuracy of 96\% with Jensen-Shannon Distance  for classifying single victim qubit. We implemented similar attack framework and statistical approach with experiments performed on VTT Q50 and achieved an accuracy of 71.68\%.  Our extended approach evaluates three scenarios (`0', `1' and no measurement) and the accuracy of the SVM model for classifying single victim qubit was 76.30\%. We also adopted this approach to analyze data across two calibration windows. The accuracy of the model in classifying single victim qubit across different calibrations was 73.89\%. For two victim qubits, we categorized the data across six configurations and the accuracy of the classifier was 53.37\%. All of our experiments were performed in VTT Q50 QC. Future research should examine similar vulnerabilities in other QC hardware like ion-trap and photonic. Such cross-platform analysis is particularly important in understanding how hardware-specific characteristics influence the security of QCs.

\section{Conclusion}
\label{conclusion}
    Measurement of a qubit can impact the readout of neighboring qubits. We analyzed how information leakage to neighboring qubit can be used to identify the measurement states as well of the number of qubits measured. The SVM classifiers trained on the experimentally derived dataset achieved an accuracy of 76.30\% for classifying one victim qubit and 53.37\% for classifying two victim qubits.

\section*{Acknowledgments}
We acknowledge the financial support of the Finnish Ministry of Education and Culture through the Quantum Doctoral Education Pilot Program (QDOC VN/3137/2024-OKM-4) and the Research Council of Finland through the Finnish Quantum Flagship project (UO 359177). The experiments in this paper have been performed on VTT’s quantum computer VTT Q50. We wish to acknowledge CSC – IT Center for Science, Finland, for their computational resources.

\printbibliography
\end{document}